\documentclass{PoS}
\usepackage{graphicx}

\def\hi {H\,{\sc i}}

\def\kms{km\,s$^{-1}$}

\def\deg{\hbox{$^\circ$}}
\def\arcmin{\hbox{$^\prime$}}

\def\fdg{\hbox{$.\!\!^\circ$}}
\def\farcm{\hbox{$.\mkern-4mu^\prime$}}

\title{RFI Mitigation for the Parkes Galactic All-Sky Survey (GASS)}

\ShortTitle{RFI Mitigation for the Parkes Galactic All-Sky Survey (GASS)}

\author{Peter Kalberla\\
        Argelander-Institut f\"ur Astronomie, Auf dem H\"ugel 71, D-53121
        Bonn, Germany\\
        E-mail: \email{pkalberla@astro.uni-bonn.de}}

      \abstract{The GASS is a survey of Galactic atomic hydrogen (HI) emission
        in the southern sky observed with the Parkes 64-m Radio
        Telescope\thanks{The Parkes Radio Telescope is part of the Australia
          Telescope which is funded by the Commonwealth of Australia for
          operation as a National Facility managed by CSIRO}~. With a
        sensitivity of 60 mK for a channel width of 1 km/s the GASS is the
        most sensitive and most accurate survey of the Galactic HI emission in
        the southern sky. We discuss RFI mitigation strategies that have been
        applied during the data reduction. Most of the RFI could be cleaned by
        using prior information on the HI distribution as well as statistical
        methods based on median filtering.  Narrow line RFI spikes have been
        flagged during the first steps of the data processing. Most of these
        lines were found to be constant over long periods of time, such data
        were replaced by interpolating profiles from the Leiden/Argentine/Bonn
        (LAB) survey \cite{GASS1}. Remaining RFI was searched for at any position by a
        statistical comparison of all observations within a distance of
        0\fdg1. The median and mean of the line emission was calculated. In
        cases of significant deviations between both it was checked in
        addition whether the associated rms fluctuations exceeded the typical
        scatter by a factor of 3. Such data were replaced by the median,  
        which is found to be least biased by RFI and other artifacts. The
        median estimator was found to be equivalent to the mean, which was
        obtained after rejecting outliers. }

\FullConference{RFI mitigation workshop\\
		 29-31 March 2010\\
		 Groningen, the Netherlands}

\begin{document}

\section{Survey description}

The Parkes Galactic All-Sky Survey (GASS) is a survey of Galactic atomic
hydrogen (\hi) emission in the southern sky for declinations $\delta <  
1\deg$ observed with the Parkes 64-m Radio Telescope. The first data release
was published by McClure-Griffiths \cite{GASS1}. The GASS is the most
sensitive, highest angular resolution large scale survey of Galactic \hi~
emission ever made in the southern sky. The intrinsic angular resolution of
the data is $14\farcm4$ (FWHM). The velocity resolution is 1.0 \kms~ and the
useful bandpass covers a velocity range $ {\rm |v_{lsr}|} < 468$ \kms. 

\section{Data reduction }
Observations and initial steps in the data reduction are described by
McClure-Griffiths \cite{GASS1}.  For the second data release (\cite{GASS2})
the stray radiation from the antenna diagram was calculated and subtracted
after an appropriate calibration of the observations. This correction may
affect profile wings, it is therefore mandatory to apply the correction for
instrumental baselines after the stray radiation was eliminated. LAB data have
been used for an initial estimate of the emission-free part of the
profile. The instrumental baseline was fitted iteratively by a 9$^{th}$ order
polynomial. Channels flagged previously as suspicious for RFI were disregarded
in this step.

After the initial steps in the data reduction all currently known systematical
influences of the instrumentation should have been taken into account. The
data base has to be self-consistent. Three main criteria for
consistency are available: 1) the derived \hi~ distribution needs to match
previous observations (e.g. the LAB data), 2) except for absorption caused by
strong continuum sources no negative signals should exist, and 3) data
observed at different seasons should not deviate from each other. Remaining
inconsistencies have in the first instance to be addressed to RFI. In the
following we describe our search for such problems and corrections that have
been applied to the data base.

\section{RFI detection }

\subsection{Flags set by {\it  Livedata}}

\begin{figure*}[!h]
   \centering
   \includegraphics[angle=-90,width=7.5cm]{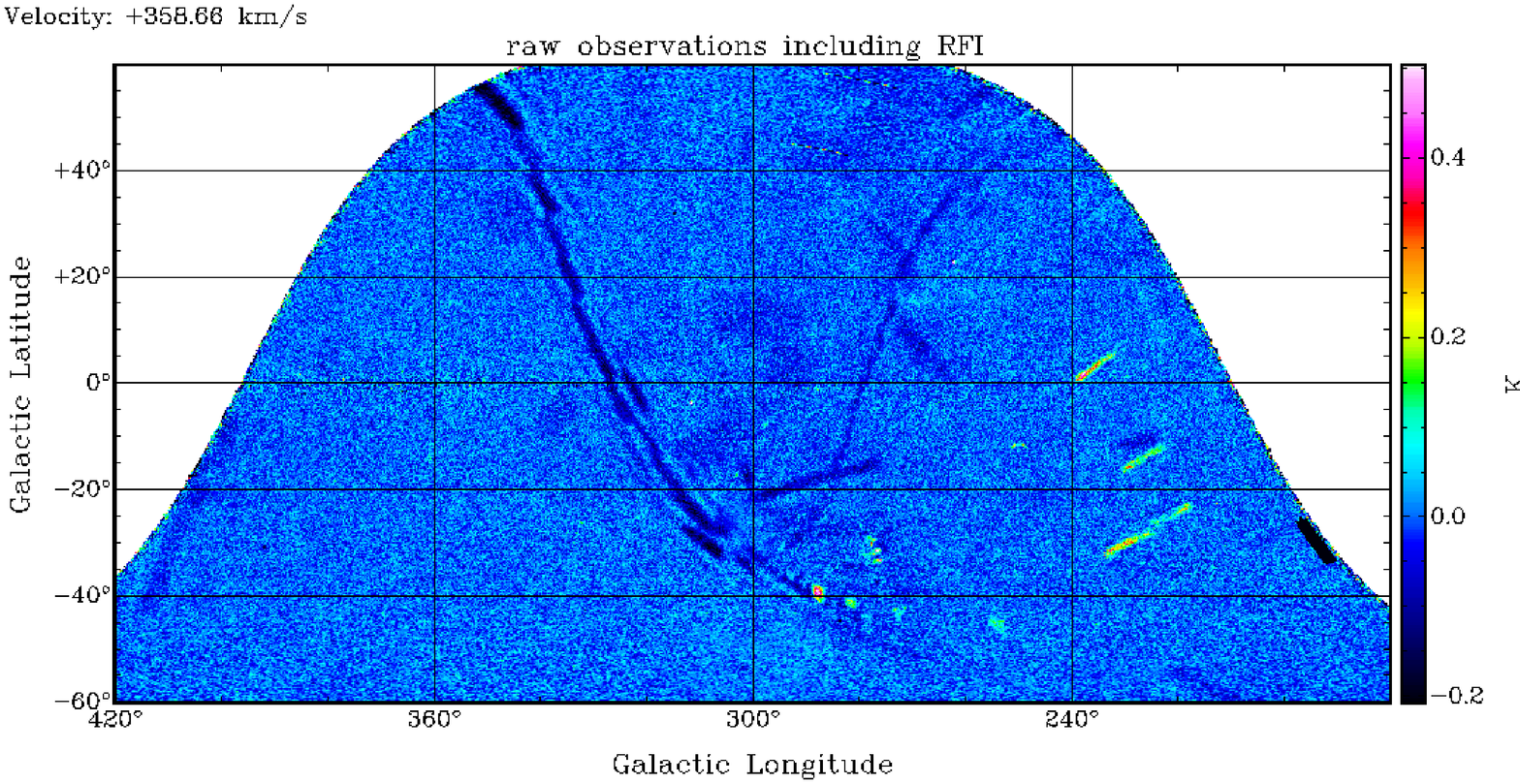}
   \includegraphics[angle=-90,width=7.5cm]{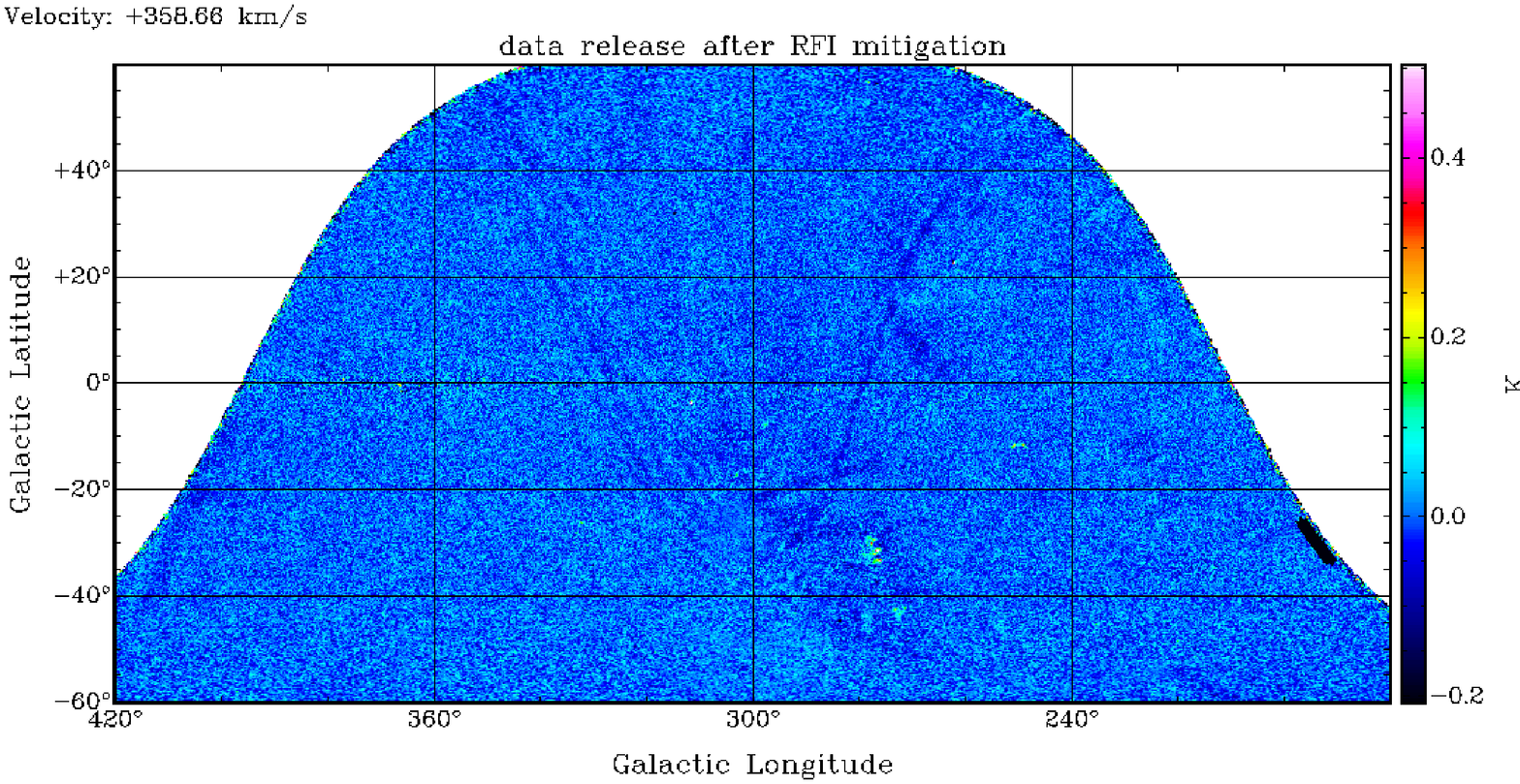}
   \caption{Typical example for extended interference patterns that were
     flagged by {\it Livedata} in the first stage of the data reduction. Left:
     as observed, right: after RFI removal using LAB data. The intensity scale
     is linear for $ -0.2 < T < 0.5$ K. Some weak structures remain but are
     mostly at $|T| < 40$ mK }
   \label{Fig_RFI1}
\end{figure*}

Easily detectable strong narrow-line interference was flagged during the basic
data reduction with {\it Livedata} \cite{GASS1}. This kind of RFI occurs
typically at fixed topocentric frequencies. As the telescope scans across the
sky the RFI moves in successive channels in the LSRK frame. Most of these
features extend over large angular distances and affect fields that have been
observed independently at different seasons (Fig. \ref{Fig_RFI1}).  This
behavior implies that the RFI remained approximately constant for long periods
in time, affecting most of the data.

This kind of strong RFI has been eliminated by replacing the flagged signal
by interpolated data from the LAB survey. Alternatively we have tried to
interpolate the observed GASS data by using the nearest ten good data
points. Using LAB data resulted in all cases in much cleaner maps. About 0.1\%
of all data have been flagged by {\it Livedata} and were replaced by
interpolated data in this first step. In addition, all profiles with more then
30 flagged channels were completely discarded. Figure \ref{Fig_RFI1} shows an
example of this kind of extended RFI (left) in comparison with the cleaned
image (right). Most of the RFI is outside the band, causing therefore a negative signal. 

\subsection{Median filter RFI excision}

\begin{figure*}[!h]
   \centering
   \includegraphics[angle=-90,width=7.5cm]{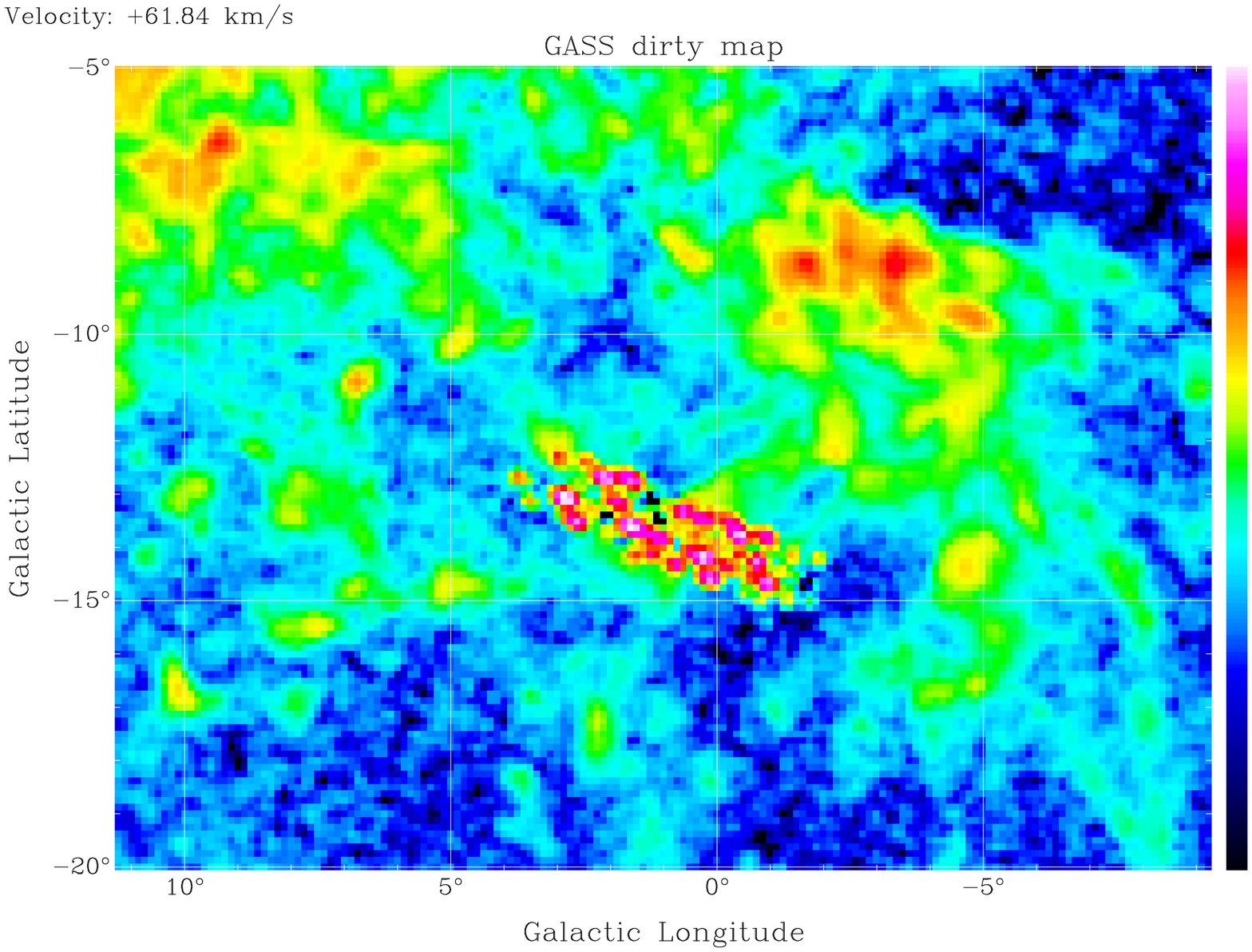}
   \includegraphics[angle=-90,width=7.5cm]{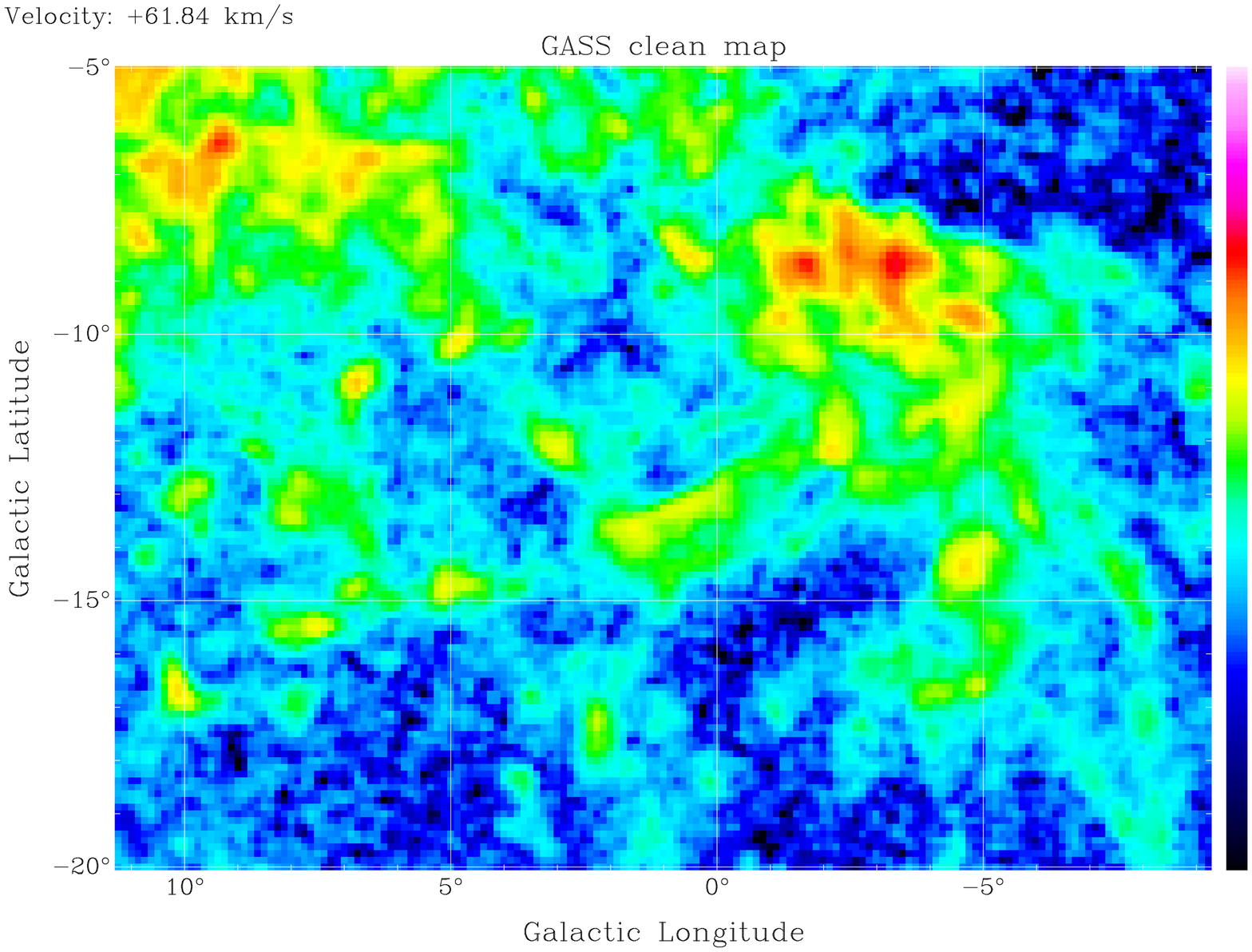}
   \caption{HI emission at ${\rm v_{lsr}} = 61.8 $ \kms.  Data heavily affected by
     RFI in scanning direction of the telescope (left) are compared with the
     clean map (right). The intensity scale for the range $ -.04 < T < 20$ K
     is logarithmic, emphasizing low level emission. Yellow colors correspond
     to a brightness temperature of $\sim 1$ K.}
   \label{Fig_RFI}
\end{figure*}

Even after removal of the channels flagged in the initial data reduction,
numerous instances of RFI remain in the data. Some appear as point sources,
but the most obvious show ``footprints'' of the 13 beam system that move on
the sky in scanning direction (Fig. \ref{Fig_RFI}). As before, the motion in
successive velocity channels is caused by changing LSR projections of the
fixed topocentric RFI frequencies. 

To search for RFI we used for each observed position a statistical comparison
of all data obtained within an angle $\epsilon = 6\arcmin$. This angle allows a
comparison of typically 40 individual spectra, a number that is sufficiently
robust for a statistical analysis without degrading the spatial resolution too
much. Using such a sample we calculate the mean for each channel, the
corresponding standard deviation, and the median, disregarding data flagged
either by {\it Livedata} or our own post-processing. The average rms noise
level $T_{RMS} = \sigma (T)$ and its one sigma deviation $\sigma (T_{RMS})$
from the average was determined for the low level emission part of the profile
($T_B < 0.5$K).

We have two tests for RFI.  In the first, channels with rms deviations $\Delta
T_{RMS} > 3 \sigma (T_{RMS})$ were considered as highly suspicious for RFI
contamination and investigated in more detail. We use as a second indication
of RFI circumstances where the mean and median differ from each other by more
than one standard deviation.

To eliminate outliers for the channels that were selected in both of the two
tests, we determined if individually observed brightness temperatures deviate
from the median by more than two times the standard deviation. In the case of
a purely random distribution this would exclude 5\% of the good data, and we
consider this an acceptable cost to eliminate true RFI.  We flag outliers and
repeat, calculating new mean and median estimators and once more excluding any
outliers.

Our two sigma limit may be compared with the more rigorous method for the
elimination of suspect data proposed by \cite{Peirce1852}, which is based on
probability theory. Accordingly a rejection of a data point that deviates from
the mean at a two sigma level is justified if there are 13 independent
observed data values. For $\epsilon = 6\arcmin$ we have on average
40 individual profiles and a two sigma limit is appropriate if about 10\% of
the data are suspect. This condition is valid for many cases but we find also
situations where 20\% of the measurements are outliers. In this case a cutoff
at $1.6~ \sigma(T_{RMS})$ would be more appropriate according to Peirce 
\cite{Peirce1852}. We deviate from Peirce's criterion in two ways; to
simplify the calculations we use a fixed $2~ \sigma(T_{RMS})$ cutoff and we
define outliers by their deviations from the median and not, as originally
proposed by Peirce, from the mean. The latter is essential since outliers
caused by RFI can be found at ten to hundred times the standard deviation, a
circumstance that was not considered by Peirce; in such cases the mean is
usually seriously affected.

To replace values outside the two sigma range, the best available estimate is
the median value scaled by the proper beam weighting function, for details we
refer to the extended discussion by \cite{Barnes2001}. We
replace the previously selected data points by the weighted median and set a
flag to distinguish medians from observed values. A fraction $\sim 5 \cdot
10^{-4}$ of the data is affected by RFI and replaced by the weighted median.

One of the criteria that we used to detect RFI was the fact that mean and
median should not differ by more than one standard deviation. We applied this
criterion also to test whether the median estimator is consistent with the
mean of all data after exclusion of the outliers. Both methods were found to
produce nearly identical results. An example of RFI in scanning direction and
the result after application of our median filer is given in
Fig. \ref{Fig_RFI}.

\begin{figure*}[!t]
   \centering
   \includegraphics[angle=-90,width=7.5cm]{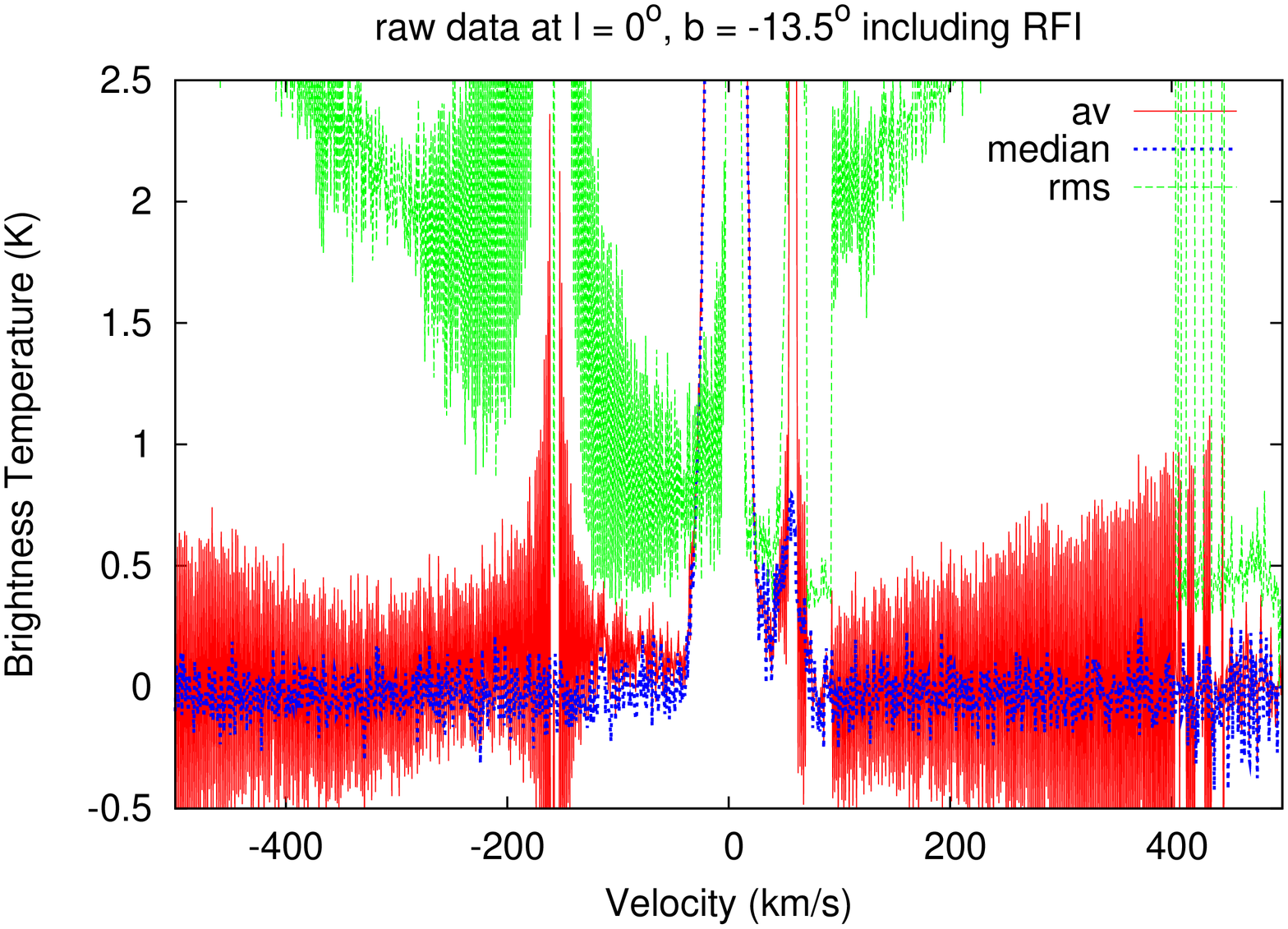}
   \includegraphics[angle=-90,width=7.5cm]{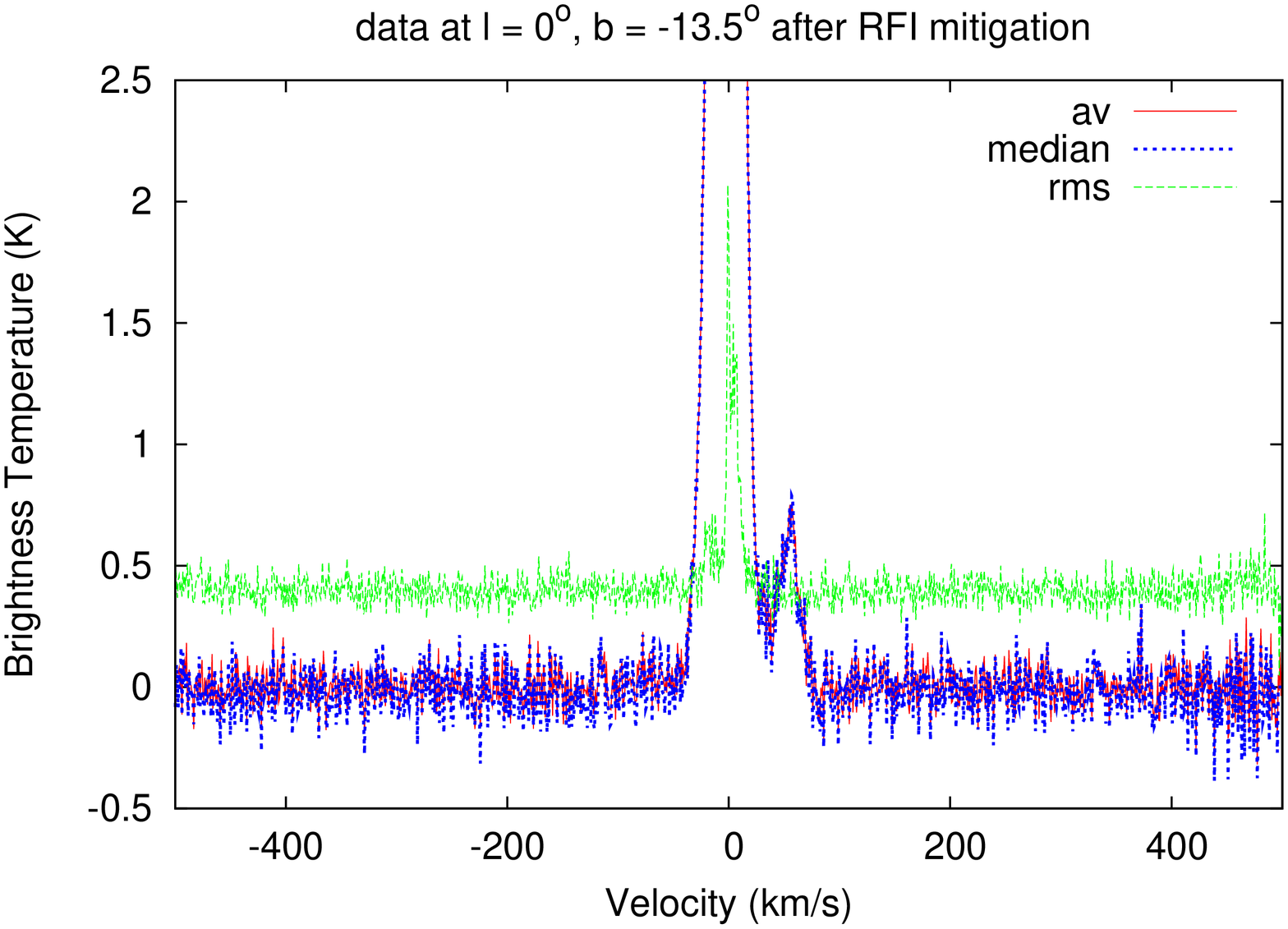}
   \caption{Data at the position $l = 0\deg$, $b = -13\fdg5$ before and after
     RFI mitigation. The average is plotted in red, the rms scatter in green,
     and the median in blue. }
   \label{Fig_RFIprof}
\end{figure*}

Figure \ref{Fig_RFIprof} shows spectral details for one position in this
field. Displayed are the mean (red), the rms scatter (green), and the median
(blue) derived for all profiles that are closer than $\epsilon < 6\arcmin$ to
the position $l = 0\deg$, $b = -13\fdg5$. Data in the left plot suffer from
strong interference at ${\rm v_{lsr}} \sim 157$ \kms, $55 < {\rm v_{lsr}}
< 58$ \kms, and at $ {\rm v_{lsr}} > 400$ \kms. Only a part of the data
was flagged by {\it Livedata} and replaced by values from the LAB. Despite all
defects the median (blue) is essentially unaffected by RFI. The data after RFI
elimination are shown in the plot at the right side. Profiles for mean and
median are within the noise identical, the rms is close to the expected value
of 0.4 K and shows no strong enhancements. We find some fluctuations close to
the line emission at ${\rm v_{lsr}} \sim 0$ \kms. This may have been caused by
genuine fluctuations in the \hi~ line emission. To avoid a possible
degradation of such fluctuations, no automatic RFI cleaning was applied for
$T_B > 0.5$K. These two plots demonstrate how greatly the rms is affected by
RFI, and how deviations between median and mean are obvious. Our RFI filter is
triggered by such deviations.

\subsection{RFI  in emission line regions}

RFI that falls in channels containing \hi~ emission was usually removed
successfully by the above procedure, except in regions with strong spectral
gradients. Examples of such critical cases are positions with strong
absorption lines. Sources with a continuum flux exceeding 200 mJy were
excluded from RFI rejection.

To avoid any possible degradation of the signal due to an automatic RFI
filtering, the median filters described previously were restricted
to data with low brightness temperatures $T_{max} < 0.5 $K. The
remaining data were checked visually. Only a few cases showed some residual
RFI. For these we increased the filter limits such that only channels with
sufficient large rms deviation $\Delta T_{RMS}$ were affected. We verified
that the remaining data were unaffected by the median filter. As a general
rule we tried to avoid as far as possible any modification of the \hi~
emission lines through the RFI filtering process. This implies, however, that
some RFI may remain undetected and be present in the final data.

Some of the spectra were affected so severely by RFI that they needed to be
removed completely. We eliminated all profiles whose mean rms within the
baseline region exceeded the average noise level at its position by a factor
of three. About 0.3\% of the observed profiles were rejected for this reason.

\section{Results}

For HIPASS, as well as for the first data release of the GASS, imaging was
performed by {\it Gridzilla} using weighted median estimation which is robust
against RFI and other artifacts. For the second data release, we use separate
and independent procedures for RFI rejection and imaging.  As described in the
previous section, our median filter is largely consistent with the HIPASS
median filter algorithm for a radius $\epsilon = 6\arcmin$. However, the major
difference is that we apply this filter only to individual channels if two
conditions apply: 1) an exceedingly large scatter is observed with a 3 times
larger dispersion than the mean dispersion caused by thermal noise and 2) mean
and median differ by more than one standard deviation.  Only data that deviate
from the median by $\Delta T_{RMS} > 2 \cdot \sigma (T_{RMS})$ are replaced by
the median estimate. 

A blind search for RFI at any given position demands quite elaborate
calculations. In total $10^9$ profiles need to be read in, some of them have
to be updated after RFI excision. The original SDFITS database does not allow
an efficient access to individual profiles. To optimize the I/O operations all
data have been converted into a new database that allows fast random
access. After this the cleaning process took several weeks.

All data have been inspected visually, only in very few cases the automated
cleaning procedure failed. The algorithm is based on a robust determination of
a clean median. This implies that less then 50\% of the data need to be
unaffected by RFI. Except for the time-independent RFI that was replaced by
LAB data, this conditions was obviously met. The data were observed in two
independent coverages with two different settings for the local oscillators.
For such a setup time variable RFI would affect only 25\% of the data,
redundancy allows in this case to excise such data without degrading the date
significantly.

We found that in total somewhat less than 0.5\% of all data was affected by
RFI and needed to be eliminated. This amount appears to be low but one should
take into account that the 21 cm line of the Galactic neutral hydrogen is
observed in the protected band. To demonstrate how dramatic the data are
degraded by RFI we provide on our web server a movie that shows the RFI that
was eliminated from the GASS database. Also data can be downloaded from
http://www.astro.uni-bonn.de/hisurvey/gass/.

{\bf ACKNOWLEDGMENTS:} This project was supported by Deutsche
Forschungsgemeinschaft, grant KA1265/5 and the participation in the conference
by a travel support of the WP3 Engineering Forum of RadioNet-FP7.

\end{document}